\documentclass[prb,twocolumn]{revtex4}

\usepackage{amssymb}
\usepackage{graphicx}
\usepackage{dcolumn}
\usepackage{bm}

\begin{document}

\title{Non--equilibrium Anisotropic Phases, Nucleation and Critical Behavior
in a Driven Lennard--Jones Fluid}
\author{J. Marro, P.L. Garrido, and M. D{\'i}ez--Minguito}
\affiliation{Institute `Carlos I' for Theoretical and Computational Physics, and
Departamento de Electromagnetismo y F\'{\i}sica de la Materia, Universidad
de Granada, E-18071 - Granada, Spain.}
\date{\today}

\begin{abstract}
We describe short--time kinetic and steady--state properties of the
non--equilibrium phases, namely, solid, liquid and gas anisotropic phases in
a driven Lennard--Jones fluid. This is a computationally--convenient two-dimensional model
which exhibits a net current and striped structures at low temperature, thus
resembling many situations in nature. We here focus on both critical
behavior and details of the nucleation process. In spite of the anisotropy
of the late--time \textquotedblleft spinodal
decomposition\textquotedblright\ process, earlier nucleation seems to
proceed by \textit{Smoluchowski coagulation} and \textit{Ostwald ripening},
which are known to account for nucleation in equilibrium, isotropic
lattice systems and actual fluids. On the other hand, a detailed analysis of
the system critical behavior rises some intriguing questions on the
role of symmetries; this concerns the computer and field--theoretical
modeling of non-equilibrium fluids.
\end{abstract}


\maketitle

\section{\label{SecI}Introduction}
Steady states in non--equilibrium many--particle systems typically involve a
constant flux of matter, charge, or some other quantity and, consequently,
stripes or other spatial anisotropies show up at appropriate scales \cite%
{haken,garr,cross,Marro}. This occurs during segregation in driven sheared
systems \cite{exp,beysens2,follow}, flowing fluids \cite{rheology2}, shaken
granular matter \cite{reis,sanchez}, and non--equilibrium liquid--liquid
binary mixtures \cite{liqliq}, and it has been reproduced in computer
simulations of driven colloidal \cite{loewen3} and fluid \cite{Marro,hurtado}
systems, for instance. Further examples are the anisotropies observed in
both high--temperature superconductors \cite{cuprates0,cuprates1} and 
electron gases \cite{2deg1,mosfet}. The ripples shaped by the wind in sand
deserts \cite{dunes,dunes2} and the lanes and trails formed by living
organisms and vehicle traffic \cite{helbing,lanes} also share some of the
essential physics.

Lacking theory for the \textquotedblleft thermodynamic\textquotedblright\
instabilities causing the observed striped structures, one tries to link
them to the microscopic dynamics of suitable model systems. For two decades,
the \textit{driven lattice gas} (DLG) \cite{KLS}, namely, a
computationally--convenient model system in which particles diffuse under an
external driving \textquotedblleft field\textquotedblright , has been a
theoretical prototype of anisotropic behavior \cite{Marro,Zia,odor}. This
model was recently shown to be unrealistic in some essential sense, however 
\cite{manolo0}. Particle moves in the DLG are along the principal lattice
directions, and any site can hold one particle at most, so that a particle
impedes the one behind to jump freely along the direction which is favored
to model the action of the field. Consequently, the lattice geometry acts
more efficiently in the DLG as an ordering agent than the field itself,
which occurs rarely ---never so dramatically--- in actual cooperative
transport. In fact, actual situations may in principle be more closely
modeled by means of continuum models, and this peculiarity of the DLG
implies that it lacks a natural off-lattice extension \cite{manolo0}.

Here we present, and analyze numerically a novel non--equilibrium
off-lattice, Lennard--Jones (LJ) system which is a candidate to portray some
of the anisotropic behavior in nature. The model, which involves a driving
field of intensity $E,$ reduces to the celebrated (equilibrium) LJ \textit{%
fluid} \cite{smit,Allen} as $E\rightarrow 0.$ For any $E>0,$ however, it
exhibits currents and anisotropic phases as in many observations out of
equilibrium. In particular, as the DLG, our model in two dimensions shows
striped steady states below a critical point. We also observe critical
behavior consistent with the equilibrium universality class. This is rather
unexpected in view of the criticality reported both for the DLG and in a
related experiment \cite{exp}. On the other hand, concerning the early--time
relaxation before well--defined stripes form by spinodal decomposition, we
first observe ---as in previous studies of relaxation towards equilibrium---
effective diffusion of small droplets, which is followed by monatomic
diffusion probably competing with more complex processes. It is very likely that our observations here concerning nucleation, coexistence, criticality, and phases morphology hold also in a number of actual systems.

The paper is organized as follows. In section II we define the model, and
section III is devoted to the main results as follows. \S\ III.A describes
the early--time segregation process as monitored by the excess energy, which
measures the droplets surface. \S\ III.B describes some structural
properties of the steady state, namely, the radial and azimuthal distribution
functions, and the degree of anisotropy. \S\ III.C, which depicts some
transport properties, is devoted to an accurate estimate of the
liquid--vapor coexistence curve and the associated critical indexes. Section
IV contains a brief conclusion.

\section{The model}
Consider $N$ particles of equal mass (set henceforth to unity) in a $d-$%
dimensional box, $L^{d},$ with periodic boundary conditions. Interactions
are via the truncated and shifted pair potential \cite{Allen}: 
\begin{equation}
\phi (r)=\left\{ 
\begin{array}{cc}
\phi _{LJ}(r)-\phi _{LJ}(r_{c}) & \text{if }r<r_{c}\  \\ 
0 & \text{if }r\geqslant r_{c},%
\end{array}%
\right.
\end{equation}%
where 
\begin{equation}
\phi _{LJ}(r)=4\epsilon \left[ (\sigma /r)^{12}-(\sigma /r)^{6}\right]
\end{equation}%
is the LJ potential, $r$ stands for the interparticle distance, and $r_{c}$
is a \textit{cut-off} that we set $r_{c}=2.5\sigma $. The parameters $\sigma 
$ and $\epsilon $ are, respectively, the characteristic length and energy
---that we use in the following to reduce units as usual.
\begin{figure}
\includegraphics[width=6cm]{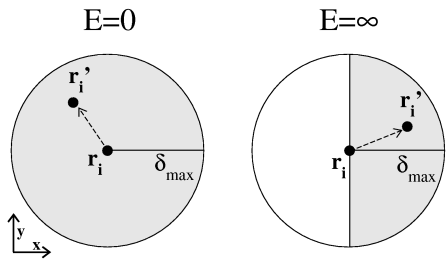}
\caption{\label{figure1} Schematic representation of the
region (grey) which is accesible to a given particle as a consequence of a
trial move for $E=0$ (left) and $E=\infty $ (right), assuming the
\textquotedblleft infinite\textquotedblright\ field points $\hat{x},$
horizontally.}
\end{figure}

Time evolution is by microscopic dynamics according to the transition
probability per unit time (\textit{rate}):%
\begin{equation}
\omega ^{\left( E\right) }\left( \mathbf{\eta }\rightarrow \mathbf{\eta }%
_{i}\right) =\chi ^{(E)}\times \min \left\{ 1,e^{-\Delta \Phi
_{i}/T}\right\} .  \label{rate}
\end{equation}%
Here, 
\begin{equation}
\chi ^{(E)}=\frac{1}{2}\left[ 1+\tanh \left( E\hat{x}\cdot \vec{\delta}%
_{i}\right) \right] ,  \label{bias}
\end{equation}%
$E$ is the intensity of a uniform external field along a principal lattice
direction, say $\hat{x},$ $\mathbf{\eta }\equiv \left\{ \vec{r}_{1},\ldots ,%
\vec{r}_{N}\right\} $ stands for any configuration of energy 
\begin{equation}
\Phi (\mathbf{\eta })=\sum_{i<j}\phi \left( \left\vert \vec{r}_{i}-\vec{r}%
_{j}\right\vert \right) ,
\end{equation}
where $\vec{r}_{i}$ is the position of particle $i$ that can be anywhere in
the $d-$torus, $\mathbf{\eta }_{i}$ equals $\mathbf{\eta }$ except for the
displacement of a single particle by $\vec{\delta}_{i}=\vec{r}_{i}^{\prime }-%
\vec{r}_{i},$ and $\Delta \Phi _{i}\equiv \Phi (\mathbf{\eta }_{i})-\Phi (%
\mathbf{\eta })$ is the cost of such displacement.

It is to be remarked that $\chi ^{(E)},$ as defined in (\ref{bias}),
contains a drive bias (see Fig.~\ref{figure1}) such that the rate (\ref{rate}%
) lacks invariance under the elementary transitions $\mathbf{\eta
\leftrightarrows \eta }_{i}.$ Consequently, unlike in equilibrium, there is
no detailed balance for toroidal boundary conditions if $E>0.$

We report here on the results from a series of Monte Carlo (MC) simulations
using a neighbor--list algorithm \cite{Allen}. Simulations concern fixed
values of $N,$ with $N\leq 10^{4},$ particle density $\rho =N/L^{d}$ within
the range $\rho \in \left[ 0.2,0.6\right] ,$ and temperature $T\in \left[
10^{-2},10^{5}\right] .$ Following the fact that most studies of striped
structures, e.g., many of the ones mentioned in the first paragraph of
section \ref{SecI}, concern two dimensions ---in particular, the DLG
critical behavior is only known with some confidence for $d=2$ \cite%
{Marro,beta4,beta5}--- we restricted ourselves to a two dimensional torus.
The maximum particle displacement is $\delta _{\text{max}}=0.5$ in our
simulations. We report below on steady--state averages over 10$^{6}$
configurations, and kinetic or time averages over 40 or more independent
runs.
\begin{figure}
\includegraphics[width=7.2cm]{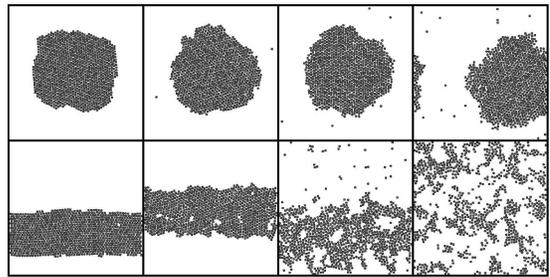}
\caption{\label{figure2} Typical steady--state
configurations for $E=0$ (top row) and $E\rightarrow \infty $ (bottom row)
at $T=$ 0.10, 0.15, 0.30 and 0.35, respectively, from left to right. This is
for $N=1000$ and $\protect\rho =0.30.$}
\end{figure}

The distribution of displacements $\vec{\delta}_{i}$ is uniform, except that
the new particle position $\vec{r}_{i}^{\prime }$ is (most often in our
simulations) sampled only from within the half--forward semi--circle of
radius $\delta _{\text{max}}$ centered at $\vec{r}_{i},$ as illustrated in
the right graph of Fig.~\ref{figure1}. This is because the \textit{%
infinite--field} limit, $E\rightarrow \infty ,$ turns out to be most
relevant, and this means, in practice, that any displacement contrary to the
field is forbidden. This choice eliminates from the analysis one parameter
and, more importantly, it happens to match a physically relevant case. As a
matter of fact, simulations reveal that any external field $E>0$ induces a
flux of particles along $\hat{x}$ ---which crosses the system with toroidal
boundary conditions--- that monotonically increases with $E$, and eventually
saturates to a maximum. This is a realistic stationary condition in which
the thermal bath absorbs the excess of energy dissipated by the drive.
\begin{figure}
\includegraphics[width=7.2cm]{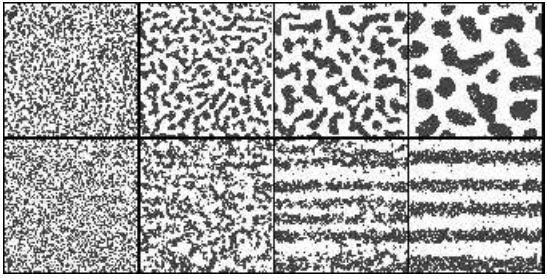}
\caption{\label{figure3} Typical configurations for $E=0$
(top row) and $E\rightarrow \infty $ (bottom row) as time proceeds during
relaxation from a disordered state (as for $T\rightarrow \infty )$ at $t=0$.
The graphs are, respectively from left to right, for times $t=$ 10$^{2},$ 10$%
^{4},$ 10$^{5},$ and 10$^{6}$ MC steps. This is for $N=7500$, $\protect\rho %
=0.35,$ and $T=0.275,$ below the corresponding transition temperature.}
\end{figure}

\section{Main results}
Fig.~\ref{figure2} illustrates late--time configurations, i.e., the ones
that typically characterize the steady state, as the temperature $T$ is
varied. These graphs already suggest that the system undergoes an
order--disorder phase transition at some temperature $T_{E}.$ This happens
to be of second order for any $E>0,$ as in the equilibrium case $E=0.$ We
also observe that $T_{E}$ decreases monotonically with increasing $E,$ and
that it reaches a well--defined minimum, $T_{\infty },$ as $E\rightarrow
\infty .$

Fig.~\ref{figure2} also shows that, at low enough temperature, an
anisotropic interface forms between the condensed phase and its vapor; this
extends along $\hat{x}$ throughout the system at intermediate densities.

\subsection{Phase segregation kinetics}
Skipping microscopic details, the kinetics of phase segregation at late
times looks qualitatively similar to the one in other non--equilibrium
cases, including driven lattice systems \cite{hurtado} and both
molecular--dynamic \cite{zeng} and Cahn--Hilliard \cite{ludo} representations of sheared fluids, while it essentially differs from the one in the
corresponding equilibrium system. This is illustrated in Fig.~\ref{figure3}.
One observes, in particular, condensation of many stripes ---as in the graph
for $t=10^{5}$ in Fig.~\ref{figure3}--- into a single one ---as in the first
three graphs at the bottom row in Fig.~\ref{figure2}. This process
corresponds to an anisotropic version of the so--called \textit{spinodal
decomposition }\cite{spinod}, which is mainly characterized by a tendency
towards minimizing the interface surface as well as by the existence of a
unique relevant length, e.g., the stripe width \cite{hurtado}. A detailed
analysis of this late regime, which has already been studied for both equilibrium \cite%
{marro2,bray} and non--equilibrium cases, including the DLG \cite%
{hurtado,levine}, will be the subject of a separate report.
\begin{figure}
\includegraphics[width=7.2cm]{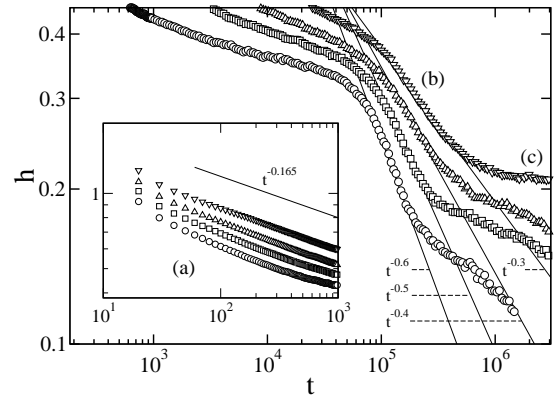}
\caption{\label{figure4} Time evolution of the enthalpy
per particle for $N=7500,$ $\protect\rho =0.35$ and, from top to bottom, $%
T=0.200,$ 0.225, 0.250 and 0.275. Straight lines are a guide to the eye; the
slope of each line is indicated. The inset shows the detail at early times.
(For clarity of presentation, the main graph includes a rescaling of the
time corresponding to the data for $T=0.250,$ 0.225 and 0.200 by factors 2,
3 and 3, respectively.)}
\end{figure}

Detailed descriptions of early non--equilibrium nucleation are rare as
compared to studies of the segregation process near completion. Following an
instantaneous quench from a disordered state into $T<T_{\infty }\left( \rho
\right) ,$ one observes in our case that small clusters form, and then some
grow at the expenses of the smaller ones but rather independently of the
growth of other clusters of comparable size. This corresponds to times $%
t<10^{5}$ in Fig.~\ref{figure3}, i.e., before many well--defined stripes
form. We monitored in this regime the excess energy or enthalpy, $H\left(
t\right) ,$ measured as the difference between the averaged internal energy
at time $t>0$ and its stationary value. This reflects more accurately the
growth of the condensed droplets than its size or radius, which are
difficult to be estimated during the early stages \cite{toral,chinos}.
Furthermore, $H\left( t\right) $ may be determined in microcalorimetric
experiments \cite{marro3}.

The time development of the enthalpy density $h\left( t\right) =H\left(
t\right) /N$ is depicted in Fig.~\ref{figure4}. This reveals some
well--defined regimes at early times.

The first regime, (a) in the inset of Fig.~\ref{figure4}, follows a power
law $t^{-\theta }$ with $\theta \approx 0.165$ ---which corresponds to the
line shown in the graph--- independently of the temperature investigated.
This is the behavior predicted by the \textit{Smoluchowski coagulation} or
effective cluster diffusion \cite{binder2}. The same behavior was observed
in computer simulations for $E=0$ and also reported to hold in actual
experiments on binary mixtures \cite{toral,marro3}. This suggests the early
dominance of a rather stochastic mechanism, in which the small clusters
rapidly nucleate, which is practically independent of the field, i.e., it is not
affected in practice by the drive. The indication of some temperature
dependence in equilibrium \cite{chinos}, which is not evident here, might
correspond to the distinction between \textit{deep} and \textit{shallow
quenches} made in Ref.~\cite{toral} that we have not investigated out of
equilibrium.

At latter times, there is a second regime, (b) in Fig.~\ref{figure4}, in
which the anisotropic clusters merge into filaments and, finally, stripes.
We observe in this regime that $\theta $ varies between 0.3 and 0.6 with
increasing $T$. \textit{Ostwald ripening }\cite{lifshitz}, consisting of
monomers diffusion, predicts $\theta =1/3.$ It is likely that regime (b)
describes a crossover from a situation which is dominated by monomers at low
enough temperature to the emergence of other mechanisms \cite{bray,baum}
which might be competing as $T$ is increased.

Finally, one observes a regime, (c) in Fig.~\ref{figure4}, which
corresponds to the beginning of spinodal decomposition.
\begin{figure}
\includegraphics[width=7.2cm]{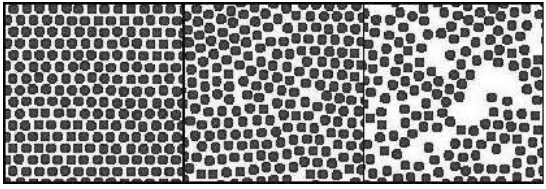}
\caption{\label{figure5} Details of
the structure in the low$-T,$ solid phase as obtained by zooming in into
configurations such as the ones in Fig.~\ref{figure2}. This is for $%
T=0.05,$ $0.12$ and 0.25, from left to right, respectively.}
\end{figure}

\subsection{Structure of the steady state}
For any $E>0,$ the anisotropic condensate changes from a solid--like
hexagonal packing of particles at low temperature (e.g., $T=0.10$ in Fig.~\ref%
{figure2}), to a polycrystalline or perhaps glass--like structure with
domains which show a varied morphology at (e.g.) $T=0.12.$ The latter phase
further transforms, with increasing temperature, into a fluid--like structure
at (e.g.) $T=0.30$ and, finally, into a disordered, gaseous state.

More specifically, the typical situation we observe at low temperature is
illustrated in Fig.~\ref{figure5}. At sufficiently low temperature, $%
T=0.05$ in the example, the whole condensed phase orders according to a
perfect hexagon with one of its main directions along the field direction $%
\hat{x}.$ This is observed in approximately 90\% of the configurations that
we generated at $T=0.05,$ while all the hexagon axis are slanted with
respect to $\hat{x}$ in the other 10\% cases. As the system is heated up,
the stripe looks still solid at $T=0.12$ but, as illustrated by the second
graph in Fig.~\ref{figure5}, one observes in this case several coexisting
hexagonal domains with different orientations. The separation between
domains is by line defects and/or vacancies. Interesting enough, as it will
be shown later on, both the system energy and the particle current are
practically independent of temperature up to, say $T=0.12.$ The hexagonal
ordering finally disappears in the third graph of Fig.~\ref{figure5},
which is for $T=0.25;$ this case corresponds to a fluid phase according to
the criterion below.
\begin{figure}
\includegraphics[width=7.2cm]{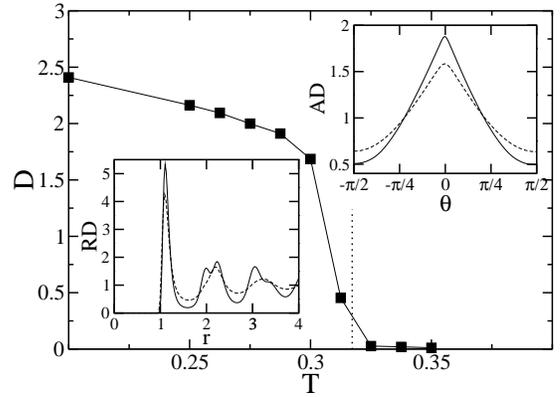}
\caption{\label{figure6} Data from simulations for $%
N=7000$ and $\protect\rho =0.35$. The main graph shows the degree of
anisotropy, as defined in the main text, versus temperature. The vertical
dotted line denotes the transition temperature. The lower (upper) inset
shows the radial (azimuthal) distribution at $T=0.20,$ full line, and $T=0.30,$
dashed line.}
\end{figure}

A close look to the structure is provided by the radial distribution (RD),%
\begin{equation}
g\left( r\right) =\rho ^{-2}\left\langle \sum_{i<j}\delta \left( \mathbf{r}%
_{i}-\mathbf{r}_{j}\right) \right\rangle ,
\end{equation}
i.e., the probability of finding a pair of particles a distance $r$ apart,
relative to the case of a random spatial distribution at same density. This
is shown in the lower inset of Fig.~\ref{figure6}. At fixed $T,$ the driven
fluid is less structured that its equilibrium counterpart, suggesting that
the field favors disorder. This is already evident in Fig.~\ref{figure2},
and it also follows from the fact that the critical temperature decreases
with increasing $E.$

The essential anisotropy of the problem is revealed by the azimuthal
distribution (AD) defined 
\begin{equation}
\alpha (\theta )=N^{-2}\left\langle \sum_{i<j}\delta \left( \theta -\theta
_{ij}\right) \right\rangle ,
\end{equation}%
where $\theta _{ij}\in \lbrack 0,2\pi )$ is the angle between the line
connecting particles $i$ and $j$ and the field direction $\hat{x}.$ Except
at equilibrium, where this is uniform, the AD is $\pi /2-$periodic with
maxima at $k\pi $ and minima at $k\pi /2,$ where $k$ is an integer. The AD
is depicted in the upper inset of Fig.~\ref{figure6}.

We also monitored the \textit{degree of anisotropy,} defined as the distance 
\begin{equation}
D=\int_{0}^{2\pi }\left\vert \alpha -1\right\vert ,  \label{funcD}
\end{equation}%
which measures the deviation from the equilibrium, isotropic case, for which 
$\alpha (\theta )=1,$ independent of $\theta .$ The function (\ref{funcD}),
which is depicted in the main graph of Fig.~\ref{figure6}, reveals the
existence of anisotropy even above the transition temperature. This shows
the persistence of non--trivial two--point correlations at high temperatures
which has been demonstrated for other non--equilibrium models \cite{pedro}.
\begin{figure}
\includegraphics[width=7.2cm]{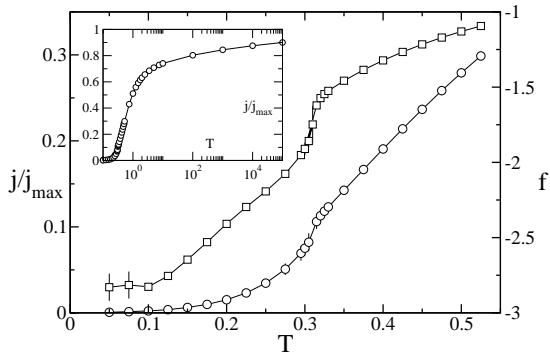}
\caption{\label{figure7} Temperature dependence of the
mean energy (squares; the scale is on the right axis) and normalized net
current (circles; scale at the left) for $N=5000$ and $\protect\rho =0.30$
under \textquotedblleft infinite\textquotedblright\ field. The inset shows
the $T-$dependence of the current over a wider range.}
\end{figure}

\subsection{Coexistence curve}
The transition points may be estimated from the temperature dependence of
the mean potential energy per particle, 
\begin{equation}
f=N^{-1}\left\langle \Phi (\mathbf{\eta })\right\rangle ,  \label{ener}
\end{equation}%
and from the net current $j,$ defined as the mean displacement per MC step
per particle. Fig.~\ref{figure7} shows well--defined changes of slope in
both magnitudes when the phase transforms from solid to liquid ($T\approx
0.12$) and then to disorder ($T\approx 0.30$). The persistence of
correlations is again revealed by the fact that the current is nonzero for
any, even low $T,$ though it is small, and roughly independent of $T$, in
the solid phase. The energy (\ref{ener}) behaves linearly with temperature
for $T\in \left( 0.12,0.3\right) ,$ as expected for a fluid phase. The
maximum value of the current, $j_{max}=4\delta _{max}/3\pi ,$ is only
reached for $T\rightarrow \infty $. The way this limit is approached is
illustrated in the inset of Fig.~\ref{figure7} where the grow is shown to be slower than
exponential.

\begin{figure}
\includegraphics[width=7.2cm]{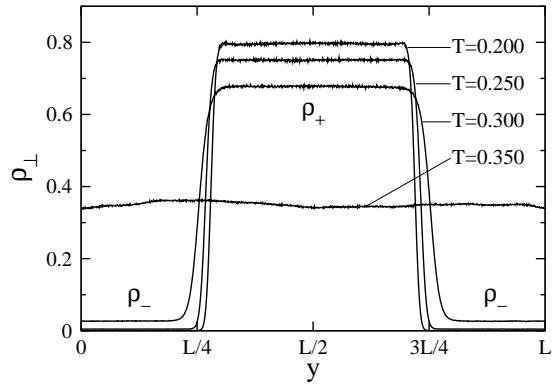}
\caption{\label{figure8} Density profiles transverse to
the field for $N=7000,$ $\protect\rho =0.35,$ and different temperature, as
indicated. The coexisting densities, $\protect\rho _{\pm },$ are indicated.}
\end{figure}

A main issue concerning the steady state is the liquid--vapor coexistence
curve and the associated critical behavior. The (non--equilibrium)
coexistence curve may be determined from the density profile transverse to
the field. This is illustrated in Fig.~\ref{figure8}.

At high enough temperature ---in fact, already at $T=0.35$ in this case for
which the transition temperature is slightly above 0.3--- the local density
is roughly constant around the mean system density, $\rho =0.35$ in Fig.~\ref%
{figure8}. As $T$ is lowered, the profile accurately describes the existence
of a single stripe of condensed phase of density $\rho _{+}$ which coexists
with its vapor of density $\rho _{-}.$ The interface becomes thinner and
smother, and $\rho _{+}$ increases while $\rho _{-}$ decreases, as $T$ is
decreased.

\begin{figure}
\includegraphics[width=7.2cm]{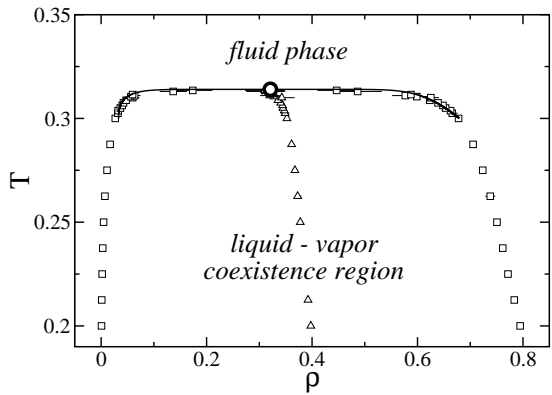}
\caption{\label{figure9} Coexistence curve (squares) for
the LJ non-equilibrium model obtained from the density profiles in Fig.~\ref{figure8}. The fluid phase and the coexistence region are
indicated. The triangles are the arithmetic mean points, which serve to
compute the critical parameters. The large circle at the top of the curve
locates the critical point, and the solid line is a fit using the Wegner
expansion and the rectilinear diameter law with the critical parameters
given in Table I.}
\end{figure}

As in equilibrium, one may use $\rho _{+}-\rho _{-}$ as an order parameter.
The result of plotting $\rho _{+}$ and $\rho _{-}$ at each temperature
results in the non-symmetric liquid-vapor coexistence curve shown in Fig.~%
\ref{figure9}. The same result follows from the current, which in fact
varies strongly correlated with the local density. Notice that the accuracy of our estimate
of $\rho _{\pm }$ is favored by the existence of a linear interface. This is
remarkable because we can therefore get closer to the critical point than in
equilibrium. Furthermore, we found that the rectilinear diameter law, 
\begin{equation}
\frac{1}{2}(\rho _{+}+\rho _{-})=\rho _{\infty }+b_{0}(T_{\infty }-T),
\end{equation}%
and the scaling law (the first term of a Wegner-type expansion \cite{wegner}%
),%
\begin{equation}
\rho _{+}-\rho _{-}=a_{0}(T_{\infty }-T)^{\beta },
\end{equation}%
can be used here to estimate the critical parameters with higher accuracy
than in the equilibrium case \cite{papa}. The simulation data in Fig.~\ref%
{figure9} then yields the values in Table I, which are confirmed by the
familiar log--log plots. Compared to the equilibrium critical temperature
reported by Smit and Frenkel \cite{smit}, one has that $T_{0}/T_{\infty
}\approx 1.46,$ i.e., the change is opposite to the one for the DLG \cite%
{Marro}. This confirms the observation above that the field acts in the
non--equilibrium LJ system favoring disorder.

The fact that the order--parameter critical exponent is relatively small may
already be guessed by noticing the extremely flat coexistence curve in Fig.~\ref{figure9}. This is similar to the corresponding curve for the
equilibrium two--dimensional LJ fluids \cite{smit,33,34}, and it is fully
consistent with the equilibrium Onsager value, $\beta =1/8.$ We therefore
believe that our model belongs to the Ising universality class. In any case,
one may discard with confidence both the DLG value $\beta \approx 1/3$ as
well as the mean field value $\beta =1/2$ which was reported for fluids
under shear \cite{exp} ---both cases would produce a hump visible to the
naked eye in a plot such as the one in Fig.~\ref{figure9}. One may argue
that this result is counterintuitive, as our model apparently has the
short--range interactions and symmetries that are believed to characterize
the DLG.

\section{Conclusion}
In summary, the present (non--equilibrium) two-dimensional Lennard--Jones system, in which
particles are subject to a constant driving field, has two main general
features. On one hand, this case is more convenient for computational
purposses, than others such as, for instance, standard molecular--dynamics
realizations of driven fluid systems. On the other, it seems to contain the
necessary essential physics to be useful as a prototypical model for
anisotropic behavior in nature.

This model reduces to the familiar LJ case for zero field. Otherwise, it
exhibits some arresting behavior, including currents and striped patterns.
We have identified two processes which seem to dominate early nucleation
before anisotropic spinodal decomposition sets in. Interesting enough, they
seem to be identical to the ones characterizing a similar situation in
equilibrium.

We have also concluded that the model critical behavior is consistent with
the Ising one for $d=2$ but not with the critical behavior of the driven
lattice gas. This is puzzling. For instance, using the language of
statistical field theory, symmetries seem to bring our system closer to the
non--equilibrium lattice model than to the corresponding equilibrium case.
The additional freedom of the present, off--lattice system, which in
particular implies that the particle--hole symmetry is violated ---which
induces the coexistence--curve asymmetry in Fig.~\ref{figure9} in accordance
with actual systems--- are likely to matter more than suggested by some
naive intuition.

Further study of the present non--equilibrium LJ system and its possible
variations is suggested. A principal issue to be investigated is the
apparent fact that the full non--equilibrium situations of interest can be
described by some rather straightforward extension of equilibrium theory.
We here report on some indications of this concerning early nucleation and
properties of the coexistence curve. No doubt it would be interesting to
compare more systematically the behavior of models against the varied
phenomenology which was already reported for anisotropic fluids. This should also help
a better understanding of non--equilibrium critical phenomena.

We acknowledge very useful discussions with M. A. Mu\~{n}oz and F. de los
Santos, and financial support from MEyC and FEDER (project FIS2005-00791).

\begin{center}
$%
\begin{tabular}{l|l|l}
$\rho _{\infty }$ & $T_{\infty }$ & $\beta $ \\ \hline\hline
0.321(5) & 0.314(1) & 0.10(8)%
\end{tabular}%
\medskip $

{\small TABLE I: Critical indexes\quad }
\end{center}


\begin{thebibliography}{99}
\bibitem{haken} H. Haken, Rev. Mod. Phys. \textbf{47}, 67 (1975).
\bibitem{garr} L. Garrido, ed., \textit{\textquotedblleft Far from Equilibrium Phase Transitions\textquotedblright }, Springer--Verlag, Berlin 1989.
\bibitem{cross} M. C. Cross and P. C. Hohenberg, Rev. Mod. Phys. \textbf{65}, 851 (1993)
\bibitem{Marro} J. Marro and R. Dickman, \textit{\textquotedblleft Nonequilibrium Phase Transitions in Lattice Models\textquotedblright } , Cambridge University Press, Cambridge 1999.
\bibitem{exp} D. Beysens and M. Gbadamassi, Phys. Rev. A \textbf{22}, 2250 (1980).
\bibitem{beysens2} C.K. Chan, F. Perrot, and D. Beysens, Phys. Rev. A \textbf{43}, 1826 (1991).
\bibitem{follow} 
A. Onuki, J. Phys.: Condens. Matter \textbf{9}, 6119 (1997).
\bibitem{rheology2} R. G. Larson, \textit{\textquotedblleft The Structure and Rheology of Complex Fluids\textquotedblright } , Oxford University Press, New York 1999.
\bibitem{reis} P. M. Reis and T. Mullin, Phys. Rev. Lett. \textbf{89}, 244301 (2002).
\bibitem{sanchez} P. S\'{a}nchez, M. R. Swift, and P. J. King, Phys. Rev. Lett. \textbf{93}, 184302 (2004).
\bibitem{liqliq} C. K. Chan, Phys. Rev. Lett. \textbf{72}, 2915 (1994).
\bibitem{loewen3} J. Dzubiella, G. P. Hoffmann, and H. L\"{o}wen, Phys. Rev. E \textbf{65}, 021402 (2002).
\bibitem{hurtado} P. I. Hurtado, J. Marro, P. L. Garrido, and E. V. Albano, Phys. Rev. B \textbf{67}, 014206 (2003).
\bibitem{cuprates0} J. Hoffman, E. W. Hudson, K. M. Lang, V. Madhavan, H. Eisaki, S. Uchida, and J. C. Davis, Science, \textbf{295}, 466 (2002).
\bibitem{cuprates1} J. Strempfer, I. Zegkinoglou, U. R\"{u}tt, M.v. Zimmermann, C. Bernhard, C. T. Lin, Th. Wolf, and B. Keimer, Phys. Rev. Lett. \textbf{93}, 157007 (2004).
\bibitem{2deg1} U. Zeitler, H.W. Schumacher, A.G.M. Jansen, R.J. Haug, Phys. Rev. Lett. \textbf{86}, 866 (2001).
\bibitem{mosfet} B. Spivak, Phys. Rev. B \textbf{67}, 125205 (2003).
\bibitem{dunes} H. Yizhaq, N. J. Balmforth, and A. Provenzale, Physica D \textbf{195}, 207 (2004).
\bibitem{dunes2} B. Andreotti, Ph. Claudin, and O. Pouliquen, arXiv:cond-mat/0506758.
\bibitem{helbing} D.Helbing, \textit{Rev. Mod. Phys.} \textbf{73}, 1067 (2001).
\bibitem{lanes} D. Chowdhury, K. Nishinari, and A. Schadschneider, Phase Trans. \textbf{77}, 601 (2004).
\bibitem{KLS} S.Katz, J. L. Lebowitz, and H. Spohn, J. Stat. Phys. \textbf{34}, 497 (1984).
\bibitem{Zia} B. Schmittmann and R. K. P. Zia, in \textit{\textquotedblleft Phase Transitions and Critical Phenomena\textquotedblright}, Vol. 17, Academic, London 1996.
\bibitem{odor} G. \'{O}dor, Rev. Mod. Phys. \textbf{76}, 663 (2004).
\bibitem{manolo0} M. D\'{\i}ez--Minguito, P. L. Garrido, and J. Marro, Phys. Rev. E \textbf{72}, 026103 (2005).
\bibitem{smit} B. Smit and D. Frenkel, J. Chem. Phys. \textbf{94}, 5663 (1991).
\bibitem{Allen} M. Allen and D. Tidlesley, \textit{\textquotedblleft Computer Simulations of Liquids\textquotedblright} , Oxford University
Press, Oxford 1987.
\bibitem{beta4} A. Achahbar, P. L. Garrido, J. Marro, and M. A. Mu\~{n}oz, Phys. Rev. Lett. \textbf{87}, 195702 (2001).
\bibitem{beta5} E. V. Albano and G. Saracco, Phys. Rev. Lett. \textbf{88}, 145701 (2002); \textit{ibid} \textbf{92}, 029602 (2004).
\bibitem{zeng} R. Yamamoto and X.C. Zeng, Phys. Rev. E \textbf{59}, 3223 (1999).
\bibitem{ludo} L. Berthier, Phys. Rev. E \textbf{63}, 051503 (2001).
\bibitem{spinod} K. Binder and P. Fratzl, in \textquotedblleft Phase Transformations in Materials\textquotedblright , G. Kostorz ed., Wiley-VCH Verlag 2001.
\bibitem{marro2} J. Marro, J. L. Lebowitz, and M. H. Kalos, Phys. Rev. Lett. \textbf{43}, 282 (1979).
\bibitem{bray} A. Bray, Adv. Phys. \textbf{43}, 357 (1994).
\bibitem{levine} E. Levine, Y. Kafri, and D. Mukamel, Phys. Rev. E \textbf{64}, 026105 (2001).
\bibitem{toral} R. Toral and J. Marro, Phys. Rev. Lett. \textbf{54}, 1424 (1985).
\bibitem{chinos} S. Y. Huang, X. W. Zou, and Z. Z. Jin, J. Phys.: Condens. Matter \textbf{13}, 7343 (2001).
\bibitem{marro3} J. Marro, R. Toral, and A. M. Zahra, J. Phys. C \textbf{18}, 1377 (1985).
\bibitem{binder2} K. Binder and D. Stauffer, Phys. Rev. Lett. \textbf{33}, 1006 (1974).
\bibitem{lifshitz} W. Ostwald, Z. Phys. Chem. \textbf{37}, 385 (1901); I. Lifshitz and V. Slyozov, J. Phys. Chem. Solids \textbf{19}, 35 (1961); C. Wagner, Z. Elektrochem. \textbf{65}, 58 (1961).
\bibitem{baum} T. Baumberger, F. Perrot, and D. Beysens, Phys. Rev. A \textbf{46}, 7636 (1992).
\bibitem{pedro} P. L. Garrido, J. L. Lebowitz, C. Maes, and H. Spohn, Phys. Rev. A \textbf{42}, 1954 (1990).
\bibitem{wegner} F. Wegner, Phys. Rev. B \textbf{5}, 4529 (1972).
\bibitem{papa} This is in spite of the fact that these fits, and the
so--called \textit{MC Gibbs method}, which are widely used for fluids in
thermal equilibrium because of its accuracy when estimating
coexistence--curve properties \cite{Pana0}, have no justification out of
equilibrium.
\bibitem{Pana0} A.Z. Panagiotopopoulos, \textit{Molecular Phys. }\textbf{61}, 813 (1987).
\bibitem{33} R.R. Singh, K. S. Pitzer, J. J. de Pablo, and J. M. Pravsnitz, J. Chem. Phys. \textbf{92}, 5463 (1990).
\bibitem{34} A.Z. Panagiotopopoulos, Int. J. Thermophys. \textbf{15}, 1057 (1994)
\end{thebibliography}
\end{document}